\documentstyle[12pt]{article}
\newcommand{\be}{\begin{equation}}
\newcommand{\ee}{\end{equation}}
\newcommand{\ba}{\begin{eqnarray}}
\newcommand{\ea}{\end{eqnarray}}

\newcommand{\no}{\nonumber \\}

\begin{document}
\begin{titlepage}
\pagestyle{empty}
\vspace{1.0in}
\begin{flushright}
March 1997
\end{flushright}
\vspace{1.0in}
\begin{center}
\begin{large}
{\bf{BARYON DENSITY AND}}\\
{\bf THE DILATED CHIRAL QUARK MODEL}\\
\end{large}
\vskip 1.0in
Youngman Kim and Hyun Kyu Lee\\
\vskip 0.5in
{\small {\it Department of Physics, Hanyang University\\
Seoul, Korea}}
\end{center}
\vspace{2cm}
\begin{abstract}
We calculate perturbatively the effect of density on hadronic properties
using the chiral quark model implemented by the QCD trace anomaly
to see the possibility of constructing Lorentz invariant Lagrangian at finite
density.
We calculate the density dependent masses of the
constituent quark, the scalar field
and the pion in one-loop order using the technique of thermo field
dynamics. In the chiral limit, the pion remains massless at finite density.
It is found that the tadpole type corrections lead to the decreasing
masses with increasing baryon density, while the radiative corrections induce
Lorentz-symmetry-breaking terms.
We found in the large $N_c$ limit with large scalar mass that
the tadpoles dominate and the mean-field approximation is reliable, giving
rise a Lorentz-invariant Lagrangian with masses decreasing as the baryon
density increases.

\end{abstract}
\hspace{.4in}PACS numbers 21.65.+f, 25.75.+r, 12.39. Fe Ki, 24.85.+p
\end{titlepage}

\section {\it {Introduction}}
\indent

It has been discussed by a number of authors
 that QCD trace anomaly plays
an important
role at finite density and temperature.
Dilaton fields which mimic the
QCD trace anomaly are used to scale non-linear $\sigma$-model to
study phase transitions\cite{ellis}\cite{Rho}
and  Nambu-Jona-Lasinio model to study quark and gluon condensate\cite{ripka}
at finite density.
A relativistic hadronic model with dilaton fields for nuclear matter
and finite nuclei is constructed and applied to the one-baryon loop level to
finite nuclei\cite{tang}.
It has been also suggested by Brown and Rho \cite{Rho} that the scale
anomaly of QCD suitably incorporated into an effective chiral Lagrangian
allows one to study dense-medium effects in a simple mean-field
approximation provided the parameters of the Lagrangian scaled in medium.
The predicted scaling law (called ``BR scaling")
has been found to be consistent with a variety of nuclear phenomena
\cite{Brown,LKB,KOSEF}.
Starting  with chiral quark
model of constituent quarks and pions that incorporates the QCD
scale anomaly, Beane and van Kolck\cite{Beane}
showed that the  ``mended symmetry" of Weinberg \cite{weinberg}
can be  realized
in the dilaton limit with the effective Lagrangian.
  As discussed in \cite{Brown}, there is a close
relation between
the physics of BR scaling and that of the dilated chiral quark model
of Beane and van Kolck \cite{Beane}.
Thus the dilated chiral quark model
is supposed to describe in some qualitative and heuristic way the state of matter
near the chiral phase
transition much as BR scaling is to describe the ``dropping" hadron masses
near the phase transition \cite{BBR}. In the recent work\cite{klr},
this aspect of the dilated chiral quark model
 was exploited to  predict the temperature dependence of
hadronic properties near the phase transition: this calculation
confirms what has always been tacitly assumed, namely,
a more rapid decreasing of $f_{\pi}, m_{\sigma},$ and $m$
with temperature in the presence of matter density compared with the
zero-density
$\sigma$ model.
In these developments, the question as to how relevant the dilaton
limit is to dense nuclear matter was not, however, properly addressed:
for example, how the Lorentz-invariant description of dilaton limit in the
presence of Fermi sea could be a good approximation.

In this paper, we make an attempt to answer the question by
calculating density effects starting
from the chiral Lagrangian implemented with the QCD trace
anomaly.
We calculate perturbatively
the effect of density  on hadronic properties using the chiral quark model
with dilaton fields
 from which Bean and van Kolck derive, in the dilaton limit,
the dilated chiral quark model.
Since we are interested mostly in density effects, we take the low-temperature
limit and focus on the parts that explicitly
depend on the density.
Using large $N_c$ approximation\cite{coleman},
 we show that the terms which break Lorentz
invariance can be neglected at large $N_c$ and for large dilaton mass
and discuss the physical relevance of Lorenz-invariant Lagrangian at
finite density.

The paper is structured as follows.
In section 2, we present the chiral quark model
with the dilaton fields and
analyze  the one-loop structures of the model using large $N_c$ approximation.
We calculate the density-dependent masses of the
pion ($m_{\pi}$),
the constituent quark ($m$) and
the scalar field ($m_{\chi}$) in section 3 at one-loop order using
the technique of thermo field
dynamics.
In section 4, we discuss the relevance of mean-field approximation in our
perturbative calculations and summarize the results.
The basics of the thermo field dynamics are
summarized in Appendix I.
Some details of our calculations are given in Appendix II.

\section{{\it Effective Chiral Lagrangian}}
\indent

The effective chiral-quark Lagrangian\cite{mg} implemented with the QCD
conformal anomaly is\cite{Beane}
\ba
L &=& \bar\psi i(\not\! \partial + \not\! V ) \psi + g_{A}
\bar \psi\not\! A \gamma_{5}\psi
-  \frac{m}{f_d} \bar\psi \psi\chi +\frac{1}{4} \frac{f_\pi^2}{f_d^2}
tr(\partial_{\mu}U \partial^{\mu} U^{\dag})\chi^2\no
 &&+ \frac{1}{2} \partial_{\mu}\chi\partial^{\mu}\chi
- V(\chi)+... \label{sicq}
\ea
where
$V_\mu = \frac{1}{2}(\xi^\dagger
\partial_\mu\xi + \xi\partial_\mu\xi^\dagger)$,
$A_\mu = \frac{1}{2}i(\xi^\dagger
\partial_\mu\xi - \xi\partial_\mu\xi^\dagger)$ with $\xi^2 =U=
\exp(\frac{i2\pi_iT_i}{f_\pi})$ and $g_A=0.752$ is the
axial-vector coupling constant for the constituent quark
\footnote{Since the role of gluons in the chiral quark
model is negligible, we will ignore the terms containing gluons\cite{mg}
\cite{keaton}.}.
The potential term for the dilaton fields,
 \footnote{The scalar
field that figures in the dilaton limit must be the quarkonium component
that enters in the trace anomaly, not the gluonium component that remains
``stiff" against the chiral phase transition. See \cite{Brown} for
a discussion on this point.}
\ba
V(\chi) = -\frac{m_{\chi}^2}{8f_d^2 }[\frac{1}{2} \chi^4 -
\chi^4 \ln(\frac{\chi^2}{f_d^2})],
\ea
reproduces the trace anomaly of QCD at the effective
Lagrangian level.
We assume that this potential chooses the ``vacuum" of the broken chiral and
scale symmetries, $<0|\chi|0>=f_d$ and
the dilaton mass is determined by
 $m_\chi^2 = \frac{\partial^2V(\chi)}{\partial\chi^2}|_{\chi= f_d}$.

After shifting the field, $\chi\rightarrow f_d +\chi^\prime$, eq.(\ref{sicq}) becomes
\ba
L &=& \bar\psi i(\not\! \partial + \not\! V ) \psi + g_{A}
\bar \psi\not\! A \gamma_{5}\psi -m \bar\psi \psi
- \frac{m}{f_d} \bar\psi \psi\chi^\prime\no
&+&\frac{1}{4}f_\pi^2
tr(\partial_{\mu}U \partial^{\mu} U^{\dag})
+\frac{1}{2} \frac{f_\pi^2}{f_d}
tr(\partial_{\mu}U \partial^{\mu} U^{\dag})\chi^\prime
+\frac{1}{4} \frac{f_\pi^2}{f_d^2}
tr(\partial_{\mu}U \partial^{\mu} U^{\dag})\chi^{\prime 2}\no
&+& \frac{1}{2} \partial_{\mu}\chi^\prime\partial^{\mu}\chi^\prime
-V(f_d + \chi^\prime)+ ...~ .\label{asf}
\ea

Expanding $U$ in terms of pion fields and collecting the
terms relevant for one-loop corrections, the Lagrangian for the fluctuating
field can be written as
\ba
L &=& \bar\psi i\not \partial \psi
 -m \bar\psi \psi
- \frac{m}{f_d} \bar\psi \psi\chi^\prime
-\frac{g_A}{f_\pi}\bar \psi\not\! \partial\pi\gamma_{5}\psi
-\frac{1}{2f_\pi^2}\epsilon_{abc}T_c
\bar \psi\not\! \partial\pi^a\pi^b\psi \no
&+&\frac{1}{2}\partial_{\mu}\pi\partial^{\mu}\pi
+\frac{f_\pi^2}{f_d}\partial_\mu\pi\partial^\mu\pi\chi^\prime
 +\frac{1}{2f_d^2}\partial_{\mu}\pi\partial^{\mu}\pi \chi^{\prime 2}\no
&+& \frac{1}{2} \partial_{\mu}\chi^\prime\partial^{\mu}\chi^\prime
-\frac{1}{2}m_\chi^2\chi^{\prime 2} -\frac{5}{6}\frac{m_\chi^2}{f_d^2}
\chi^{\prime 3} +...~.
\label{asb}
\ea

We analyze the Lagrangian using the large $N_c$
approximation at one-loop order.
From 't Hooft and Witten\cite{coleman}, we know the $N_c$ dependence of the
properties of mesons and baryons.
Some of them, which are relevant to us, are
summarized as follows:
Three meson vertiex is of order  $\frac{1}{\sqrt{N_c}}$.
Baryon-meson-baryon vertex is of order $\sqrt{N_c}$.
But, since the constituent quarks in eq.(\ref{sicq}) have color indices,
there should be some modification to the meson-fermion vertex.
The mesons are created and annihilated by quark bilinears,
\be
B=N_c^{-\frac{1}{2}}\bar\psi^a\psi^a\label{mes}
\ee
If we consider matrix element of eq.(\ref{mes}) between two color-singlet
baryons,  the order is
$N_c/N_c^{\frac{1}{2}}$ since $N_c$ quarks
are involved to be annihilated by $B$.
However in our case the ground-states are the
constituent quarks with definite color and
only one quark which has same color with ground-state quarks
can contribute the matrix element; thus the
 quark-meson-quark vetrex is of order
$1/N_c^{\frac{1}{2}}$.
So quark-meson-quark vertex is of order
$\frac{1}{\sqrt{N_c}}$ while there are no changes in $N_c$ counting for
 mesonic vertices.
Now we can determine $N_c$ dependences of one-loop diagrams.
The tadpole contribution to consitutent quark mass, Fig.2, is of order
$N_c(\frac{1}{\sqrt{N_c}})^2 = 1$ and that from Fig.3 is of order
$(\frac{1}{\sqrt{N_c}})^2 =\frac{1}{N_c}$.
The two diagrams, Fig. 4 and Fig. 5,
  for dilaton mass corrections are of the same
$N_c$ order.\\

\section{{\it Perturbative calculations}}
\indent

We compute the diagrams representing mass corrections
at one-loop approximation.
At very low temperature, the Bose-Einstein distribution function
$n_B(k) =\frac{1}{e^{\beta k} -1}$ goes to zero but the Fermi-Dirac
distribution
function $n_F(p) =\frac{1}{e^{\beta (p-\mu)} +1}$ becomes $\theta (\mu-p)$.
Thus in our case, the finite-temperature and -density corrections come
from quark propagators.
In our calculations of Feynman integrals,
we shall follow the method of Niemi and Semenoff \cite{ns}.
In the case of renormalization of finite-temperature QED \cite{do}
one encounters a new
singularity like $ \frac{1}{k}\frac{1}{e^{\beta k} -1} $.
However in our case, because pion-quark vertices depend on momentum,
we do not have any additional infrared singularities at finite temperature
and density.
At finite temperature and density, the lack of explicit Lorentz
invariance causes some ambiguities in defining renormalized masses.
The standard practice is to define density-dependent (or $T$-dependent)
mass corrections as the energy of the particle at $\vec p =0$ \cite {bla}.
In our
case, this definition may not be the most suitable one because of the
Fermi blocking from the fermions inside the Fermi sphere. We find
it simplest and most convenient to define the mass at $\vec p=0$.\\

\leftline{{\bf \underbar{Pion mass}}}
\indent

We first consider pion mass in baryonic matter at low temperature.
There are three density-dependent diagrams, Fig. 1, that may contribute to
pion mass corrections.
The first one, Fig. 1a, vanishes identically due to isospin symmetry.
Explicit calculation of the two diagrams in Fig. 1b gives,
for $\beta\rightarrow\infty $,
\ba
\Sigma_\pi(p^2) &=&i(\frac{g_A}{f_\pi})^2 tr\int\frac{d^4k}{(2\pi)^4}
\not \! p\gamma_5T^a(\not\! p+\not\! k +m)
\cdot\not \! p\gamma_5 T^a (\not\! k +m)\no
& &[\frac{i}{(p+k)^2-m^{2}}(-2\pi)\delta (k^2-m^2)
\sin^2\theta_{k_0}\no
& &+\frac{i}{k^2-m^{2}}(-2\pi)\delta ((k+p)^2-m^{2})
\sin^2\theta_{k_0+p_0}].\label{pm0}
\ea
With the change of variable on the second term,
$p+k\rightarrow k$, the above integral can be rewritten as
\ba
\Sigma_\pi(p^2) &=&-2(\frac{g_A}{f_\pi})^2 \int\frac{d^4k}{(2\pi)^3}
[\frac{-p\cdot k(p^2+2k\cdot p )+2m^{2}p^2}{p^2+2k\cdot p} \no
& &~~~~~~~~~~ + \frac{p\cdot k(p^2-2k\cdot p)+2m^{2}p^2}{p^2-2k\cdot p}]
\delta (k^2-m^{2})\sin^2\theta_{k_0}\label{pm1}\no
&=&-2p^2(\frac{g_A}{f_\pi})^2 \int\frac{d^4k}{(2\pi)^3}[\frac{2m^2}{p^2+2k\cdot p}
+\frac{2m^2}{p^2-2k\cdot p}]\no
&&~~~~~~\times\delta (k^2-m^{2})\sin^2\theta_{k_0}.\label{pm2}
\ea
Since the pion self-energy is proportional to $p^2$, the pole
of the pion propagator does not change. Hence the pion remains
massless as long as there is no explicit chiral symmetry breaking.
Naively we might expect that the pion may acquire a dynamical
mass due to dynamical screening from thermal particles or
particle density, which is the case  for QED \cite{sa} even with
gauge invariance. The reason for this is the derivative pion coupling
to the quark field.
In eq.(\ref{pm2}), the terms with $p\cdot k$ may give rise to terms
which are not proportional to $p^2$, which lead to the pion
mass correction after $dk$ integration. However, those terms are
proportional to $p^2 + 2k\cdot p$ or $p^2 - 2k\cdot p$ which are
cancelled by the their denominators and contribute nothing after
the $dk$ integration. This feature has been explicitly
demonstrated in eq.(\ref{pm2}).

Since the pole of the pion propagator does not change, we can take
the renormalization point at $p^2=0$. Then the integrand itself
vanishes, so there is no wave-function
renormalization for the pion in dense medium.\\

\leftline{{\bf \underbar{Quark mass}}}
\indent

At one-loop order, the quark self-energy is given by the diagrams
of Fig. 2 and Fig. 3.
The diagram Fig. 2 gives
\be
\Sigma^{(1)}_Q(p)=i(-i\frac{m}{f_d})^2\frac{i}{-m_\chi^2}\rho_s
\ee
where $\rho_s$ is the scalar density obtained from the fermionic
loop with thermal propagator,
\ba
\rho_s&=&-tr\int\frac{d^4p}{(2\pi)^4}(-2\pi)(\not\! p+m)\delta (p^2-m^2)
sin^2\theta_{p_0}\no
&=&\frac{m}{\pi^2}\theta(\mu-m)[\mu\sqrt{\mu^2 - m^{2}}
 - m^{2} ln ( \frac{\mu +  \sqrt{\mu^2 - m^{2}}  }{m})]
\ea
where $\vec p_F^2=\mu^2-m^2$.
With $I$ defined as
$\rho_s  = 4mI$, we get
\be
\Sigma^{(1)}_Q(p) = -(\frac{m}{f_d})^2\frac{4m}{m_\chi^2}I\label{qm1}
\ee
The radiative correction from the pion field in Fig. 3a is
\ba
\Sigma^{3a}_Q(p)=\frac{3}{8} (\frac{g_A}{f_\pi})^2
[(\not\! p +m)I].\label{fse}
\ea
Finally, the radiative correction due to the $\chi$-field in Fig. 3b is found
to be
\ba
\Sigma^{3b}_Q(p)
=-(\frac{m}{f_d})^2[\not\! J - \frac{m}{2m_{\chi}^2}I]
\ea
where $\not\! J$ is defined
\be
\not \! J\equiv \int\frac{d^4k}{(2\pi)^3}\frac{\not\! k}{2m^2-2p\cdot k
-m_\chi^2}\delta (k^2-m^2)\sin^2\theta_{k_0}.\label{jde1}
\ee

In sum, the self-energy of the quark with four momentum $(E, \vec p)$ can be
written in the form
\ba
\Sigma_Q(E, \vec p)=aE\gamma_0+b\vec\gamma\cdot\vec p +
c-(\frac{m}{f_d})^2\frac{4m}{m_\chi^2}I\label{gs}
\ea
where $a$, $b$, and $c$ are
\ba
a&=& \frac{3}{8}(\frac{g_A}{f_\pi})^2I - \frac{1}{E}(\frac{m}{f_d})^2J^0\no
b&=&-\frac{3}{8}(\frac{g_A}{f_\pi})^2I + (\frac{m}{f_d})^2
\frac{1}{\vec p^2}\vec J\cdot\vec p\no
c&=&\frac{3}{8}(\frac{g_A}{f_\pi})^2mI + (\frac{m}{f_d})^2\frac{m}{2m_\chi^2}I
\label{abc}
\ea
where we have used the relation \cite{do},
$\vec J\cdot \vec \gamma =\frac{\vec J\cdot\vec p
~\vec p\cdot\vec\gamma}{\vec p^2}$.
The detailed calculations
of $J^0$ and $\vec J\cdot\vec p$ are given in appendix II.

From eq.(\ref{abc}), we can see that the scalar field radiative
corrections violate the
Lorentz symmetry, that is, $ a \neq -b$, unless $\frac{1}{E}J^0 =
\frac{1}{\vec p^2} \vec J\cdot \vec p$.
On the other hand, the pion radiative corrections in $a$ and $b$ preserve
Lorentz covariance.
Since we have only $O(3)$ symmetry in the medium for the quark
propagation, we will adopt the conventional definition of mass
as a zero of the inverse propagator with zero momentum.
The inverse propagator of the quark is given by
\ba
G^{-1}(E,p) &=& \not\! p -m -\Sigma(E, \vec{p})) \\
&=& E(1-a)\gamma_0-(1+b)\vec\gamma\cdot\vec p - c-m
\ea
The mass is now defined as the energy which satisfies $det(G^{-1})
=0$  with $\vec p^2=0$,
\be
(1-a)E =c+m\label{dr}.
\ee
Since $a$ and $c$ are perturbative corrections,  eq.(\ref{dr}) can be
approximated to the leading order as
\be
E =m-(\frac{m}{f_d})^2\frac{4m}{m_\chi^2}I+c+am \label{dm}.
\ee
If we define the mass as the energy of the particle at finite
density, the quark mass becomes
\ba
m^*&=&m -(\frac{m}{f_d})^2\frac{4m}{m_\chi^2}I\no
&&+\frac{3}{4}(\frac{g_A}{f_\pi})^2mI \no
&&- (\frac{m}{f_d})^2J^0 + (\frac{m}{f_d})^2\frac{m}{2m_\chi^2}I.\label{qms}
\ea
The first line in eq.(\ref{qms}) is just the result of the mean-field
approximation, in
which only the tadpole diagram, Fig. 2, is taken into account, as
further elaborated on in the section 4.  While the
dropping of the quark mass with density is obvious in the mean-field
approximation, it is not clear whether it is still true when the
radiative corrections, Fig.3, are included as in the second and
the third lines in eq.(\ref{qms}).
Thus the result obtained at the one-loop order does not indicate
in an unambiguous way that the quark mass is scaling in medium
according to BR scaling \cite{Rho}. The specific behavior depends on
the strength of the coupling constants involved in the theory, $g_A, m, f_\pi$
and $f_d$. This does not seem to be the correct physics for
BR scaling as evidenced in Nature.
However, as discussed in section 2,
the tadpole contribution to consitutent quark mass, Fig.2, is of order
$1$ and that from Fig.3 is of order
$\frac{1}{N_c}$.
So we can neglect the Fig.3 for large $N_c$ and obtain in-medium quark mass,
\be
m^*=m -(\frac{m}{f_d})^2\frac{4m}{m_\chi^2}I.
\ee
In the large $N_c$ limit, therefore, the quark propagator
retains Lorentz covariance and the mass does decrease according to
the mean-field approximation.\\

\leftline{{\bf \underbar{Scalar mass}}}
\indent

Now consider the mass shift of the dilaton field (see Fig. 4 and
Fig. 5).
The tadpole diagram, Fig. 4, gives
\ba
\Sigma^{(2)}_\chi(p)&=&-\frac{5}{6}\frac{m_\chi^2}{f_d}
(-i\frac{m}{f_d})\rho_s(\frac{i}{-m_\chi^2})3 !\no
&=&-20(\frac{m}{f_d})^2I \label{dlm}
\ea
where the factor $3!$ comes from the topology of the diagram. This corresponds
to the mean-field approximation at one-loop order.

The analytic expression for the contribution from Fig. 5 can be
obtained in the large scalar mass approximation, $m^2<<m_\chi^2$,
\ba
\Sigma^{(5)}_\chi(p)&=&\frac{m^2}{m_\chi^2}
\frac{8}{\pi^2}(\frac{m}{f_d})^2[\frac{1}{2}\theta (\mu - m)
(\mu\sqrt{\mu^2 - m^2}
 - m^2 \ln ( \frac{\mu +  \sqrt{\mu^2 - m^2}  }{m})) \no
& &~~~~~~~~~~-\frac{E^2}{m_\chi^2}{\bf (}\frac{\mu(\mu^2-m^2)^{3/2}}{4m^2}
+\frac{\mu\sqrt{\mu^2-m^2}}{8}\no
& &~~~~~~~~~~-\frac{m^2}{8}
\ln (\frac{\mu+\sqrt{\mu^2-m^2}}{m})~{\bf )}~].\label{scm}
\ea
Note that the terms on the second and the third lines contribute only to
the {\it energy} of the dilaton field and hence break Lorentz invariance.
This contribution, (\ref{scm}), however can be neglected since it is
suppressed compaired to that of tadpole, eq.(\ref{dlm}), by
$\frac{m^2}{m_\chi^2}$.
Hence the Lorentz invariance is maintained approximately in the
large scalar mass approximation.
Now the renormalized dilaton mass becomes
\be
m_{\chi}^{*2}=m_\chi^2- 20(\frac{m}{f_d})^2 I.
\ee
The mass of the dilaton drops as density increases. This is consistent
with the tendency for the scalar to become dilatonic at large density,
providing an answer to the question posed at the beginning.

\section{{\it Discussion}}
\indent

So far, we have not taken into account the fact that
introducing  the thermal fluctuation can cause  further
shifting of the vacuum expectation value of
the dilaton field.
The vacuum expectation value of $\chi^\prime$-field in eq.(\ref{asf})
is zero, $<0|\chi^\prime|0>$=0, at zero density.
The introduction of the new ground state, $|F>$, which
 are already occupied by constituent
quarks with $E$$<$$ \it{E}_f$(fermi energy),
implies the necessity of shifting the vacuum.
Imposing the condition that all tadpole graphs
vanish on the physical vacuum,
in terms of Weinberg's notation\cite{web}, the condition is
\be
(\frac{\partial P(\chi^\prime)}{\partial \chi^\prime})_{\chi^\prime
=<F|\chi^\prime|F>}
+\tilde T=0\label{tp}
\ee
where $P(\chi^\prime )$ is a polynomial in $\chi^\prime$ and $\tilde T$ is the sum of
all tadpole graphs.
In our case, we are interested in the small change of vacuum
expectation value,
$\frac{<F|\chi^\prime|F>}{f_d} <1 $.
So we can drop higher powers of the $\chi^\prime$-field and retain
only the tadpoles that depend on density at one-loop order.
Then the eq.(\ref{tp}) reads
\be
-m_\chi^2\chi_0 -\frac{m}{f_d}\rho_s=0\label{nt}
\ee
where $\chi_0\equiv <F|\chi^\prime|F>$.
Then, we get the vacuum expectation value of the $\chi^\prime$-field.
\be
\chi_0 =-\frac{m}{m_\chi^2 f_d} \rho_s
\ee
This is equivalent to Fig. 1 without the external quark line.
With the shift of the $\chi^\prime$-field around $\chi_0$, i.e.
$\chi^\prime\rightarrow \chi_0+\tilde\chi$
in eq.(\ref{asf}) or
$\chi\rightarrow f_d + \chi_0+\tilde\chi$ in eq.(\ref{sicq}),
 the Lagrangian is modified
effectively to
\ba
L &=& \bar\psi i(\not\! \partial + \not\! V ) \psi + g_{A}
\bar \psi\not\! A \gamma_{5}\psi -\frac{m}{f_d}(f_d+ \chi_0) \bar\psi \psi
- \frac{m}{f_d} \bar\psi \psi\tilde\chi\no
&&+\frac{1}{4} \frac{f_\pi^2}{f_d^2}(f_d + \chi_0)^2
tr(\partial_{\mu}U \partial^{\mu} U^{\dag})\no
&&+ \frac{1}{2} \partial_{\mu}\tilde\chi\partial^{\mu}\tilde\chi
-V(f_d+\chi_0+\tilde\chi)+\cdots\label{lag}
\ea
where the ellipsis stands for the interaction terms including
pions and $\chi$-fields.
The Lagrangian, eq.(\ref{lag}), shows explicitly
the density dependence of $f_\pi^2$, $m_\chi^2$ and $m$:
\ba
f_\pi^{*2}&=&\frac{f_\pi^2}{f_d^2}(f_d+\chi_0)^2
=f_\pi^2(1+\frac{\chi_0}{f_d})^2\no
m_\chi^{*2} &=&\frac{\partial^2V(\chi)}
{\partial\chi^2 }\mid_{\chi =f_d+\chi_0}
=\frac{m_\chi^2}{f_d^2}(f_d+\chi_0)^2(1+3\ln \frac{f_d+\chi_0}{f_d})\no
&\simeq&m_\chi^2(1+\frac{\chi_0}{f_d})^2(1+3\frac{\chi_0}{f_d})
\simeq m_\chi^2(1+5\frac{\chi_0}{f_d})\no
m^*&=&\frac{m}{f_d}(f_d+\chi_0) =m(1+\frac{\chi_0}{f_d}).
\ea
This is the result obtained in the previous section.
Hence we can see that
the mean field approximation in the chiral quark model coupled to dilaton is
a good approximation in the large $N_c$ limit with
massive dilaton, $(\frac{m}{m_\chi})^2<<1$, and gives
\be
\frac{m^{*}}{m}=\frac{f_\pi^*}{f_\pi}\cong\frac{m_\chi^*}{m_\chi}\label{sl}
\ee
as predicted by BR scaling.
This is somewhat different from the observation of Nambu-Freund
model \cite{nf}
which consists of a matter field $\psi$ and a dilaton field $\phi$.
After spontaneous symmetry breaking the matter field and dilaton field
acquire masses.
There are universal dependences on the vacuum expectation values both for
the matter field and the dilaton field.
One can easily see that the universal dependence on the vacuum expectation
value is no longer valid
in our calculations due to the logarithmic potential from QCD trace anomaly.

In summary, we have found that the tadpole type corrections
lead to the decreasing
masses with increasing baryon density, while the radiative corrections induce
Lorentz-symmetry-breaking terms.
The pion remains massless at finite density in the chiral limit.
In the context of large $N_c$ approximation with large scalar mass,
tadpoles dominate and the mean-field approximation is reliable, giving
rise to a Lorentz-invariant Lagrangian with masses decreasing as the baryon
density increases according to eq.(\ref{sl}).
This analysis with large $N_c$ approximation gives a clue to construct
the Lorentz invariant lagrangian, {\it i.e}, dilated
chiral quark model which incoporate the mended symmetry
of Weinberg into chiral quark model, at finite density.
The dilated
chiral quark model can emerge in dense and hot medium -- if it does
at all -- only through
nonperturbative processes (e.g., large $N_c$ expansion) starting from
a chiral quark Lagrangian.\\

{\bf Acknowledgments}\\

We thank Mannque Rho for the useful discussions.
This work was supported in part by the Korea Ministry of Education
(BSRI-96-2441) and
in part by the Korea Science and Engineering Foundation under Grants No.
94-0702-04-01-3

\newpage
{\bf\it Appendix I}\\

We summarize the basic notation and definition of Thermo Field
Dynamics(TFD)
\cite{ma}\cite{ksa}.
TFD is a real time operator formalism in quantum
field theory
at finite temperature. The main feature of the TFD is that thermal
average of operator $A$ is defined as the expectation value with respect
to the temperature dependent vacuum, $|0(\beta)>$, which is introduced
through Bogoliubov
transformation.
\ba
<A> &\equiv&  Tr(A e^{-\beta (H -\mu)})/ Tr(e^{-\beta (H -\mu)})\no
&=& <0(\beta )\mid A\mid 0(\beta )>
\ea
where $\beta\equiv \frac{1}{k T}$, H is the total Halmiltonian of the system,
and $\mu$ is the chemical potential.
The propagator in TFD can be separated into two parts, i.e., the usual
Feymann part $G_F $ and density-dependent part $G_D$,
\be
G_{\alpha\beta} = G_{F\alpha\beta} +G_{D\alpha\beta}.
\ee
For fermion with mass $m$,
\ba
G_{F\alpha\beta} &=&(\not\! p +m)_{\alpha\beta}
\left( \begin{array}{cc}
\frac{1}{p^2-m^2 +i\epsilon} &      0                       \\
             0               &   \frac{1}{p^2-m^2 -i\epsilon}
\end{array} \right)\no
G_{D\alpha\beta} &=& 2\pi i\delta (p^2 -m^2)(\not\! p +m)_{\alpha\beta}
\left( \begin{array}{cc}
\sin^2\theta_{p_0}            &    \frac{1}{2}\sin 2\theta_{p_0}\\

\frac{1}{2}\sin 2\theta_{p_0}  &    -\sin^2\theta_{p_0}

\end{array} \right)
\ea
with
\ba
\cos\theta_{p_0}&=&\frac{ \theta(p_0)}{(1+e^{-x})^{1/2}} +
     \frac{\theta(-p_0)}{(1+e^{x})^{1/2}},\no
\sin\theta_{p_0}&=& \frac{e^{-x/2}\theta(p_0)}{(1+e^{-x})^{1/2}} -
                \frac{e^{x/2}\theta(-p_0)}{(1+e^{x})^{1/2}},
\ea
where $\alpha$ and $\beta$ are Dirac indices and $ x= \beta(p_0 - \mu) $.\\
The propagator of the scalar particle with mass $m_s$ is given by
\be
\Delta(p)= \Delta_F(p) + \Delta_D(p),
\ee
where
\ba
\Delta_F &=&
\left( \begin{array}{cc}
\frac{1}{p^2-m^2_s +i\epsilon} &      0                       \\
             0               &   \frac{-1}{p^2-m^2_s -i\epsilon}
\end{array} \right)\no
\Delta_D &=& -2\pi i\delta (p^2 -m^2_s)
\left( \begin{array}{cc}
\sinh^2\phi_{p_0}            &    \frac{1}{2}\sinh 2\phi_{p_0}\\

\frac{1}{2}\sinh 2\phi_{p_0}  &    \sinh^2\phi_{p_0}

\end{array} \right)
\ea
with
\ba
\cosh\phi_{p_0}&=&\frac{1}{(1-e^{-|y|})^{1/2}} ,\no
\sinh\phi_{p_0}&=& \frac{e^{-|y|/2}}{(1-e^{-|y|})^{1/2}} ,
\ea
where $ y= \beta p_0  $.\\

\newpage

{\bf\it Appendix II}\\

In this Appendix II, we evaluate various intergrals in the text.\\
{\bf 1}.
The contribution for the quark self energy from
 Fig. 3a(pion radiative corrections) is given by
\ba
-i\Sigma^{3a}_Q(p)&=&
-(\frac{g_A}{f_\pi})^2\int \frac{d^4k}{(2\pi)^4}\frac{i}{(p-k)^2}
(\not\! p -\not\! k)\gamma_5 T^a\no
& &\times (-2\pi)\delta (k^2-m^{2}) (\not\! k +m)(\not\! p -\not\! k)\gamma_5 T^a
\sin^2\theta_{k_0}\no
&=& i\frac{3}{4}(\frac{g_A}{f_\pi})^2\int \frac{d^4k}{(2\pi)^3}
\frac{-(p-k)^2\not\! k +2(p-k)\cdot k(\not\! p-\not\! k)
-m(p-k)^2}{(p-k)^2}\no
 & &\times \delta (k^2-m^2) \sin^2\theta_{k_0}\no
&=&i\frac{3}{4}(\frac{g_A}{f_\pi})^2\int \frac{d^4k}{(2\pi)^3}
[-\not\! k-(\not\! p-\not\! k)-m]\delta (k^2-m^2) \sin^2\theta_{k_0}\no
&=&-i\frac{3}{4}(\frac{g_A}{f_\pi})^2(\not\! p +m)\frac{1}{4\pi^2}\int
dk_0 \bar k \sin^2\theta_{k_0}\no
&=&-i\frac{3}{8\pi^2}(\frac{g_A}{f_\pi})^2(\not\! p +m)I
\ea
where $\bar k$ denotes $|\vec k|$. This gives eq.(\ref{fse}) in section 4.\\
{\bf 2}. We evaluate $J^0$ and $\vec J\cdot\vec p$ in eq.(\ref{abc}),
 which
are needed in the evaluation of the quark mass correction from $\chi$-field
 in Fig. 3b.
Firstly let us consider $J^0$,
\ba
J^0 &=&\int \frac{d^4k}{(2\pi)^3} \frac{k_0}{2m^2-2p\cdot k- m_\chi^2}
\delta (k^2-m^2)\sin^2\theta_{k_0}\no
&=&\int \frac{d^3\vec k}{(2\pi)^3}\frac{1}{2}
\frac{1}{2m^2-2 p\cdot k- m_\chi^2} \sin^2\theta_{k_0}\no
&=&\frac{1}{8\pi^2}\int_0^{k_F} d\bar k \bar k^2
\frac{1}{2\bar p\bar k}\ln (\frac{2\bar p\bar k+2m^2-2Ek_0-m_\chi^2}
{-2\bar p\bar k+2m^2-2Ek_0-m_\chi^2})
\ea
Note that due to the factor $sin^2\theta_{k_0}\mid_{\beta\rightarrow\infty}
\rightarrow \theta(\mu-k_0)$, $k_0$ is maximally order of chemical potential,
i.e., $k_0\sim \mu$. Assuming
the large dilaton mass, $m_\chi^2>>\mu^2>m^2(\sim E^2)$,
\ba
J^0&\cong&\int \frac{d^3\vec k}{(2\pi)^3}\frac{1}{2}
\frac{1}{2m^2-2 p\cdot k- m_\chi^2} \sin^2\theta_{k_0}\no
&=& -\frac{1}{12\pi^2}\frac{1}{m_\chi^2}(\mu^2-m^2)^{3/2}\label{j01}
\ea

Similarly,
\be
\vec J\cdot \vec p =  \int\frac{d^4k}{(2\pi )^3}
\frac{\vec p\cdot \vec k}{2m^2-2p\cdot k-m_\chi^2}\delta (k^2-m^2)
\sin^2\theta_{k_0}
\ee
can be approximated
\ba
\vec J\cdot \vec p
&\cong&\frac{1}{8\pi^2}\frac{4E^2}{m_\chi^4}\int dk_0 k_0^2\bar k
\no
&=&\frac{1}{8\pi^2}\frac{E^2}{m_\chi^4}
\theta (\mu -m)[\mu (\mu^2 -m^2)^{3/2}\no
& &~~~~+~\frac{m^2\mu\sqrt{\mu^2-m^2}}{2}-\frac{m^4}{2}
\ln (\frac{\mu+\sqrt{\mu^2-m^2}}{m})]
\ea
{\bf 3}.
The contribution from Fig. 5 for the dilaton self-energy, eq.(\ref{scm}),
 is given by
\ba
-i\Sigma_\chi^{(5)}(p^2) &=&-(\frac{m}{f_d})^2 tr\int\frac{d^4k}{(2\pi)^4}
(\not\! p +\not\! k +m)(\not\! k +m)(-2\pi)\no
& &~~~~~~~~~\times~[\frac{i}{(p+k)^2-m^{2}}\delta(k^2-m^{2})\sin^2\theta_{k_0}\no
& &~~~~~~~~~+\frac{i}{k^2-m^{2}}\delta((p+k)^2-m^{2})\sin^2\theta_{p_0+k_0}]\no
&=&8i(\frac{m}{f_d})^2\int\frac{d^4k}{(2\pi)^3}[\frac{p\cdot k+2m^2}
{p^2+2p\cdot k}+\frac{-p\cdot k+2m^{2}}{p^2-2p\cdot k}]
\delta(k^2-m^{2})\sin^2\theta_{k_0}\no
&=&8i(\frac{m}{f_d})^2\int\frac{dk_0dx}{(2\pi)^3}\bar k d\bar k^2
[\frac{Ek_0- \bar p\bar k x+2m^2}{p^2 + 2Ek_0-2\bar p\bar k x}\no
& &~~~~~~~~~+\frac{-Ek_0+\bar p\bar k x + 2m^{2}}{p^2-2Ek_0+2\bar p\bar k x}]
\delta(k^2-m^{2})\sin^2\theta_{k_0}\no
&=&-i(\frac{m}{f_d})^2\frac{1}{\pi^2}\int dk_0\bar k[2+\frac{p^2-4m^2}
{4\bar p k}\ln (\frac{p^2+2Ek_0-2\bar p\bar k}
{p^2+2Ek_0+2\bar p\bar k})\no
& &~~~~~~~~~- \frac{p^2-4m^2}{4\bar p k}
\ln (\frac{p^2-2Ek_0+2\bar p\bar k}{p^2-2Ek_0-2\bar p\bar k})]
\ea
In the large dilaton mass limit, $(\frac{m}{m_\chi})^2<<1$,
\ba
-i\Sigma_\chi^{(5)}(p^2) &\cong&-i\frac{8}{\pi^2}(\frac{m}{f_d})^2\int
dk_0 \bar k[\frac{m^2}{p^2}-
\frac{1}{3p^4}(\bar p^2\bar k^2+3E^2k_0^2)]\times\sin^2\theta_{k_0}\no
&\cong& -i\frac{8}{\pi^2}(\frac{m}{f_d})^2\int dk_0 \bar k[\frac{m^2}{p^2}
-\frac{E^2k_0^2}{p^4}]\no
&=&-i\frac{m^2}{m_\chi^2}
\frac{8}{\pi^2}(\frac{m}{f_d})^2[\frac{1}{2}\theta (\mu - m)
(\mu\sqrt{\mu^2 - m^2}
 - m^2 ln ( \frac{\mu +  \sqrt{\mu^2 - m^2}  }{m}) \no
& &~~~~~~~~~~-\frac{E^2}{m_\chi^2}(\frac{\mu(\mu^2-m^2)^{3/2}}{4m^2}
+\frac{\mu\sqrt{\mu^2-m^2}}{8}\no
& &~~~~~~~~~~-\frac{m^2}{8}
\ln (\frac{\mu+\sqrt{\mu^2-m^2}}{m})]
\ea
where $\bar p $ and $\bar k$ denote $|\vec p|$ and $|\vec k|$
respectively, and $p^2 = m_\chi^2$ is used. It is suppressed by
order of $\frac{m^2}{m_\chi^2}$ compared to the tadpole contribution.

\newpage
\begin{center}
{\large\bf Figure Captions}
\end{center}
Fig.1 Self-energy diagrams for pions.
 Solid lines represent the constituent quarks and
 dashed lines are for pions.
 D(F) represents thermal(Feynman) propagator.\\

\noindent Fig.2 Tadpole type diagram for quark.
  Solid lines represent the constituent quarks and
  wavy lines are  dilaton field respectively.\\

\noindent Fig.3 Self-energy diagram for quark.\\

\noindent Fig.4 Tadpole type diagram for dilaton field.\\

\noindent Fig.5 Self-energy diagram for dilaton field.


\begin{thebibliography}{99}
\bibitem{ellis} B.A. Campbell, J. Ellis and A. Olive, Nucl.
Phys. {\bf B345}, 57(1990)
\bibitem{Rho} G.E. Brown and M. Rho, Phys. Rev. Lett. {\bf 66}, 2720 (1991)
\bibitem{ripka} G. Ripka and M. Jaminon, Ann. of Phys. {\bf 216}, 51(1992)
\bibitem{tang} R.J. Furnstahl, H.-B. Tang and B.D. Serot, Phys. Rev. {\bf C52}, 1368(1995)
\bibitem{Brown} G.E. Brown and M. Rho, Phys. Rep. {\bf 269}, 333 (1996)
\bibitem{LKB} G.Q. Li, C.M. Ko and G.E. Brown, Phys. Rev. Lett. {\bf 75},
4007 (1996); Nucl. Phys., in press  and papers in preparation.
\bibitem{KOSEF} B. Friman and M. Rho, Phys. Repts., in press and
nucl-th/96022025; M. Rho, KOSEF Lecture, February 1996, Seoul, Korea;
``QCD vacuum changes in nuclei,"
talk at the International Symposium on Non-Nucleonic Degrees of Freedom
Detected in Nucleus, 2-5 September 1996, Osaka, Japan.
\bibitem{Beane}    S. Beane and U. van Kolck, Phys. Lett. {\bf B328}, 137 (1994)
\bibitem{weinberg} S. Weinberg, Phys. Rev. Lett. {\bf 65}, 1177 (1990)
\bibitem{BBR} G.E. Brown, M. Buballa and M. Rho, Nucl. Phys., in press.
\bibitem{klr} Y. Kim, H. K. Lee and M. Rho, Phys. Rev. {\bf C52}, R1184 (1995)
\bibitem{coleman} E. Witten, Nucl. Phys. {\bf B160}, 57(1979);
          G. 't Hooft, Nucl. Phys. {\bf B72}, 461(1974); S. Coleman,
          {\it Aspects of symmetry}, Cambridge University Press
\bibitem{mg}    A. Manohar and H. Georgi, Nucl. Phys. {\bf B234}, 189 (1984)
\bibitem{keaton}   G. L. Keaton, hep-ph/9612422
\bibitem{ns} A. J. Niemi and G. W. Semenoff, Nucl. Phys. {\bf B230[FS10]},
181 (1984)
\bibitem{do} J.F. Donoghue and B.R. Holstein, Phys. Rev.{\bf D28}, 340 (1983)
\bibitem{bla} J.-P. Blaizot, JKPS {\bf 25}, S65 (1992)
\bibitem{sa} S. S. Masood, Phys. Rev. {\bf D44}, 3943 (1991)
\bibitem{web} S. Weinberg, Phys. Rev. {\bf D7}, 2887 (1973)
\bibitem{nf}  P. G. O. Freund and Y. Nambu, Phys. Rev. {\bf 174}, 1741(1968)
\bibitem{ma} H. Matsumoto in {\it Progress in Quantum Field Theory}, ed.
           by H.  Ezawa and S. Kamefuchi (North-Holland, Amsterdam, 1986)
\bibitem{ksa} K. Saito, T. Maruyama and K. Soutome, Phys. Rev. {\bf C40},
407 (1989)
\end{thebibliography}
\end{document}